\input{aipcheck}

\documentclass[
    ,final            
  ]
  {aipproc}

\layoutstyle{6x9}

\usepackage{amsmath}
\usepackage{array}

\begin{document}

\newcommand{\be}{\begin{equation}}
\newcommand{\ee}{\end{equation}}
\newcommand{\pdf}{\mathcal{P}}
\newcommand{\data}{{\bf d}}
\newcommand{\mdl}{\mathcal{M}}
\newcommand{\lsim}{\,\raise 0.4ex\hbox{$<$}\kern -0.8em\lower 0.62ex\hbox{$\sim$}\,} \newcommand{\gsim}{\,\raise 0.4ex\hbox{$>$}\kern -0.7em\lower 0.62ex\hbox{$\sim$}\,}

\newcommand{\params}{\boldsymbol{\theta}}
\newcommand{\paramsU}{\boldsymbol{\theta}_\star}
\newcommand{\fsel}{f_{sel}}
\newcommand{\fobs}{f_{obs}}
\newcommand{\gal}{{gal}}
\newcommand{\stars}{{stars}}
\newcommand{\plan}{{planets}}
\newcommand{\mpl}{M_{\rm{Pl}}}
\newcommand{\rhol}{\rho_\Lambda}
\newcommand{\rhom}{\rho_m}
\newcommand{\rhor}{\rho_r}
\newcommand{\aeq}{a_{eq}}
\newcommand{\Heq}{\mathcal{H}_{eq}}
\newcommand{\omleq}{\Omega_{\Lambda, {eq}}} \newcommand{\Rmin}{R_{min}} \newcommand{\Nmax}{N_{max}}

\newcommand{\Tds}{T_{dS}}
\newcommand{\mean}{\boldsymbol{\mu}}
\newcommand{\like}{L}
\newcommand{\lnlike}{\mathcal{L}}
\newcommand{\ML}{^*}
\newcommand{\dr}{rm{d}}
\newcommand{\ie}{i.e.}
\newcommand{\eg}{e.g.}
\newcommand{\reion}{{re}}

\newcommand{\cd}{\cdot}
\newcommand{\cds}{\cdots}
\newcommand{\ip}{\int_0^{2\pi}}
\newcommand{\al}{\alpha}
\newcommand{\ba}{\beta}
\newcommand{\de}{\delta}
\newcommand{\De}{\Delta}
\newcommand{\ep}{\epsilon}
\newcommand{\Ga}{\Gamma}
\newcommand{\ka}{\tau}
\newcommand{\io}{\iota}
\newcommand{\La}{\Lambda}
\newcommand{\Om}{\Omega}
\newcommand{\om}{\omega}
\newcommand{\si}{\sigma}
\newcommand{\Si}{\Sigma}
\newcommand{\te}{\theta}
\newcommand{\ze}{\zeta}
\newcommand{\vth}{\ensuremath{\vartheta}}
\newcommand{\vph}{\ensuremath{\varphi}}
\newcommand{\MM}{\mbox{$\cal M$}}
\newcommand{\tr}{\mbox{tr}}
\newcommand{\hor}{\mbox{hor}}
\newcommand{\grad}{\mbox{grad}}
\newcommand{\cx}{\ensuremath{\mathbf{\nabla}}}
\newcommand{\life}{{\rm{life}}}

\title{What's the trouble with anthropic reasoning?}


\author{Roberto Trotta}{
  address={Astrophysics Department, Oxford University, Denys Wilkinson Building, Keble Road, Oxford OX1~3RH, UK} }

\author{Glenn D. Starkman}{
  address={Astrophysics Department, Oxford University, Denys Wilkinson Building, Keble Road, Oxford OX1~3RH, UK}
  ,altaddress={Department of Physics, Case Western Reserve University, Cleveland, OH~~44106-7079, USA} }

\begin{abstract}
Selection effects in cosmology are often invoked to ``explain''
why some of the fundamental constant of Nature, and in particular
the cosmological constant, take on the value they do in our
Universe. We briefly review this probabilistic ``anthropic
reasoning'' and we argue that different (equally plausible) ways
of assigning probabilities to candidate universes lead to totally
different anthropic predictions, presenting an explicit example
based on the total number of possible observations observers can
carry out. We conclude that in absence of a fundamental motivation
for selecting one weighting scheme over another the anthropic
principle cannot be used to explain the value of $\Lambda$.

\end{abstract}

\maketitle


\section{Introduction}

The existence of observers' selection effects in cosmology might
appear at first sight a mere tautology. Indeed, all of our
observations of the Universe are (implicitly) conditional on the
fact that we exist. A Universe without observers would never be
measured for the simple fact that there would be no one around to
make the observations.

We can easily conceive Universes where the laws of physics are
such as not to allow the existence of sentient life. For example,
a Universe without an arrow of time would be extremely hostile to
the emergence of an organized complexity as required by the
existence of intelligent observers, because the lack of causality
would arguably prevent any meaningful anticipation of physical
phenomena. Without going to such extremes, we might speculate
about what the Universe would look like if the numerical value of
physical constants were different from what is observed. In
particular, a large number of so--called ``anthropic
coincidences'' have been pointed out regarding our own Universe
(see e.g.~\cite{BarrowBook} and references therein): it appears
that any small deviation from highly fine--tuned values of
physical constants in our Universe would have catastrophic
consequences for the emergence of life as we know it.

There are a few different viewpoints as to what meaning we should
assign to the idea that physical constants (such as Newton's
constant or the fine structure constant) could be different from
what we observe. The traditional approach of physics has been to
explain the laws of nature from fundamental principles, such as
the existence and breaking of symmetries. In working out a final
theory of everything, one might have hoped that the structure of
physical reality would naturally emerge as the unique logically
and mathematically consistent possibility. In this case, the value
of natural constants would be uniquely determined by some deep,
underlying principle. The imminent realization of this
positivistic hope seems to have receded as string theory has
failed to deliver such a unique picture of fundamental physics.
Indeed, the string landscape, with its large number of different
vacua, points to a vast number of possible Universes, each with
different properties, for example the values of fundamental
constants. Opinions diverge as to the physical reality of such
alternative realizations of the Universe. We might think of this
ensemble as causally disconnected patches of this Universe, as
separate sub--universes (a.k.a. the multiverse) or as a
superposition of states (as in quantum cosmology). It seems to us
that this problem is akin to the (unresolved) interpretation of
measurement in quantum mechanics, where the physical reality of
the many--worlds picture remains unclear. This has lead some
distinguished scientists (such as the late E.T.~Jaynes) to cast
doubts on the completeness of modern quantum theory.

\section{Anthropic reasoning}

One of the most outstanding problems in fundamental physics is to
explain the incredibly low energy density of the vacuum compared
to the characteristic Planck energy density -- the cosmological
constant problem. Lacking an explanations from fundamental
arguments, much effort has been devoted to investigate the
possibility of understanding the cosmological constant values in
terms of selection effects. This goes usually under the name of
``anthropic principle'', introduced by Brandon
Carter~\cite{Carter} as a necessary limitation to the Copernican
principle that we do not hold a special place in the Universe. The
argument first given by Weinberg~\cite{Weinberg:1988cp_etal} is
that in realizations of the Universe with too large a value of the
cosmological constant structure formation cannot proceed, and
hence life as we know it will not emerge. Although Weinberg's
original upper limit of the value of the cosmological constant
energy density $\rho_\Lambda$ is more that 2 orders of magnitude
larger than what is observed, refined versions of this argument
claim to successfully ``predict'' $\rhol$ comparable to what is
actually observed, ie $\rhol/\mpl^4 \approx
10^{-123}$~\cite{Vilenkin:1994ua,Martel:1997vi,Garriga:1999bf}.

The rigorous translation of such selection effects in a
probabilistic statement is far from trivial. Let us focus on the
case where only the cosmological constant $\La$ is allowed to vary
(see \cite{Aguirre:2001zx,Tegmark:1997in,Tegmark:2005dy} for a
discussion of how the situation changes when more parameters are
varied). In a Bayesian language, we can write for the posterior
probability for $\La$ given that intelligent life exist,  \be
\label{bayes}  Pr(\La | \life) \propto Pr(\La)Pr(\life|\La),  \ee
where $Pr(\La)$ is the prior probability distribution function
(pdf) and $Pr(\life|\La)$ is the likelihood for ``life'' given a
certain value for $\La$, thus encapsulating the selection effects.
The proportionality constant (the ``evidence'') is independent of
$\La$ and can be ignored for the purpose of this discussion. Let
us discuss the prior and likelihood in turn.

\subsection{The prior distribution}

The whole point of the anthropic program is to go from a flat(ish)
prior pdf -- either describing insufficient knowledge (from the
Bayesian perspective) or assumptions about the frequency of
realizations of a fundamental theory (frequentist) -- to a
strongly peaked posterior, hopefully centered around the observed
value for $\Lambda$.

But to start with, how are we to assign the prior? Most of the
literature has interpreted the prior in a frequentist sense, and
calculated in various way the relative number of outcomes for a
large number of realizations. However, the very concept of
probability as a limiting frequency of outcomes, though natural
when applied to repeatable experiments, is not obviously
appropriate to describe the Universe as a whole. One way out of
the problem that we have only one Universe to study is to make use
of ergodic arguments in order to derive $Pr(\La)$
\cite{Tegmark:2005dy}. This is an approach whose validity remain
unproven. A more radical point of view is the Multiverse scenario,
according to which there is an infinite collection of, by
definition inaccessible, universes.  It is difficult to see how
vastly increasing the number of universes could help determine the
properties of the one universe we actually can observe, \ie\ our
own. This approach hardly seems economical in terms of explanatory
power. From an operational point of view, the idea of a ``random''
distribution of values for $\Lambda$ is meaningless unless a
mechanism for the generation of the different values is also
specified.

Some of these conceptual difficulties might be addressed by taking
a fully Bayesian approach to the problem, and understanding the
prior as an expression of our state of knowledge before we see the
data, in this case represented by the observation that sentient
life does exist in the Universe. Here, however, we are mainly
concerned with issue arising from the choice of the likelihood
function, to which now we turn out attention.

\subsection{The likelihood function: dependence on reference class}

In order to compute $Pr(\La|\life)$ in Eq.~\eqref{bayes}, we need
to specify exactly the meaning of ``life''. This is a fundamental
issue that all too often is glossed over, by simply using a more
readily calculated surrogate -- the physical number density of
galaxies, or more precisely the collapsed gas fraction. One then
assumes that the density of observers is proportional to this
quantity. So if ``life'' really stands for ``intelligent observers
just like us'', we need to ask ourselves what {\em exactly} counts
as an observer. Do future generations of humans count as separate
observers or not -- after all, we could pass on the information we
gathered to them. Or perhaps, the whole human civilization ought
to be counted as one single, collective observer? Do the ancient
Egyptians count as separate observers? What if past observers die
out, or if they forget previous measurements?

These considerations can be put in a more formal way by using the
concept of a {\em reference class of observers}. The reference
class contains all observers that are ``just like you'' in all
relevant respects. It is clear that the resulting posterior pdf
will depend on the chosen reference class, as shown by
\cite{Neal}. In other words, the choice of reference class is
equivalent to giving different probabilistic weights to different
realizations of the Universe (in a frequentist perspective). Here
we present a specific example of this effect -- we argue that
there are many plausible weighting factors (or reference classes)
for universes, and that the answers to questions such as the
expected value of $\Lambda$ depends enormously on the weighting.
In~\cite{ST06} we introduced a weighting scheme based on the
maximal number of allowed observations (MANO) in a universe. This
quantity is clearly relevant to the expected value of a constant,
say $\Lambda$, since a value that allows more observations to be
carried out will be measured more often. It also has the advantage
of being independent of how one defines constant time
hypersurfaces. As we show below, the resulting posterior pdf for
$\La$ is peaked arbitrarily close to 0, thus giving a completely
different result that the usual weighting by number density of
galaxies.

\section{Maximum Number of Allowed Observations}

We wish to evaluate the probability that an observer will measure
his or her universe  to have a vacuum energy density no smaller
than what we measure in our Universe. As the selection function
for observing $\Lambda$ in the different realizations we put
forward the total number of observations that observers can
potentially carry out over the entire life of that universe
(called MANO for brevity, for ``Maximum Allowed Number of
Observations''). This maximum number is the product of two factors
-- the number of observers and the maximum number of observations
that each observer can make. As argued above, there is a
fundamental difficulty in determining the total number of
observers in a given reference class, since we can neither compute
nor measure it. However, in the limit where observers are rare (in
a way we quantify below) the anthropic prediction for the
probability of observing $\Lambda$ will  be independent of the
density of observers. Below we focus on the second factor -- the
maximum number of observations that each observer can make -- and
offer some further comments about the many observers scenario in
the next section.

\subsection{Rare observers scenario}

For illustrative purposes and computability, we hold fixed all
parameters of the universe other than the vacuum energy density,
considering  flat Lemaitre-Friedmann-Robertson-Walker universes
with exactly the same matter and radiation contents and the same
fluctuations as our own at the time of matter--radiation equality.
This is a common setup in the literature. We consider only the
case $\Lambda> 0$. This restriction can only increase the
probability of observing $\Lambda$ equal to or greater than the
observed one, so we should interpret the probability we calculate
as an upper limit. We refer to \cite{ST06} for further details.

In a $\Lambda > 0$ universe, the minimum temperature at which a
system ({\it e.g.} an observer) can operate is the de Sitter
temperature $\Tds = \rhol^{1/2}/(2 \pi \mpl)$. (Refrigerated
subsystems can run cooler, but the energy consumption of the
refrigeration more than compensates). As discussed in detail in
\cite{Krauss:1999hj} and \cite{Krauss:2004jy}, the maximum energy
such an observer can collect is given by  \be  E_{\rm max} \simeq
\frac{1}{8}\frac{4\pi}{3}\left[(\eta_\infty - \eta_\star)
 a_\star\right]^3 \rho_m(a_\star)  \ee
 where $\eta_\star$ is the
value of conformal time when the observer starts collecting energy
and $a(\eta_\infty) = \infty$. The factor of $\sim1/8$ arises
because we assume that the decision to collect the energy is made
by the observer at the origin, and must then be communicated out
into space. During this time-consuming process, most ($\sim 7/8$)
of the volume currently within the apparent horizon is swept out
of it by the accelerating expansion. (There is an additional
suppression factor of $\sim1/8$ if one wishes to transport the
energy back to the central location, rather than use it {\it in
situ}) We will ignore the ${\cal O}(1)$ geometric prefactors and
focus on the functional dependence. In the ``rare observers''
scenario we assume that there is at most one observer within the
comoving volume accessible to each from the time that they first
become capable of making observations onward, otherwise there is a
cut--off to the maximum collectible energy introduced by
competition among observers (see below).

The number of thermodynamic processes (such as observations of
$\Lambda$) an observer can carry out is maximized if the observer
saves up $E_{\rm max} $ until the universe has reached the de
Sitter temperature. Thus \be \label{eq:nmax} \Nmax \leq E_{\rm
max}/k_B\Tds . \ee Following the arguments given above, we adopt
$\Nmax$ as a probabilistic weight in the selection function, and
since we have assumed a flat prior pdf in $\La$ we have that the
posterior $Pr(\La|\life) \propto \Nmax$.

The asymptotic limit for $\Nmax$ can be calculated analytically
(see \cite{ST06}), and one finds  \be \label{eq:Nmaxcrude}
 \Pr(R|\al_\star)
\propto \left\{
\begin{matrix}
  \frac{2}{3}\left(\al_0/\al_\star\right)^3 R^{-2},  \quad R \gg 1,  \\
  54 R^{-1}, \quad R \ll 1.
  \end{matrix}
  \right.
  \ee
Here we have introduced $R$ as the ratio of the value of the
cosmological constant in an hypothetical universe with respect to
the value it takes in our own,  $R \equiv \La/\La_0$. The quantity
$\al_\star$ is the value of the scale factor when smart observers
start collecting energy (normalized to matter--radiation
equality), and it is conveniently expressed as  \be
\label{eq:alphastar} \al_\star = {\al_0}\left(3R\right)^{-1/3}
\sinh \left( \ln(\sqrt{3} + 2) \sqrt{R} \tau \right)^{2/3},  \ee
where $\tau \equiv t_\star/t_0$ is the physical time until
observers smart enough to begin collecting energy arise, in units
of 13.7 Gyrs, the age at which such observers (us, or our
descendants) are known to have arisen in our Universe. However,
the normalization integral in \eqref{eq:Nmaxcrude} diverges
logarithmically, and is dominated by the minimum cut--off value,
$\Rmin$, if such exists. In the landscape scenario (see e.g.\
\cite{Vilenkin:2006qf} and references therein), for instance, the
number of vacua is estimated to be of order $10^{500}$, and
therefore the corresponding minimum value of $\Lambda$ can perhaps
be taken to be $\Lambda_{min} \sim 10^{-500}\mpl^4$, or $\Rmin
\sim 10^{-377}$. One could of course go from \eqref{eq:Nmaxcrude}
to a joint posterior for $R$ and $\tau$, but this would involve
the specification of a prior on $\tau$, thus introducing further
uncertainty in the problem, given our ignorance about when smart
observers are likely to arise. We prefer instead to evaluate the
posterior probability of $R > 1$, equivalent to the probability of
measuring $\La > \La_0$, for a few representative choices of
$\tau$. With the above choice of cut--off value, and for a few
values of $\tau=0.1, 1, 10$ we obtain a very small probability of
observing a value of $\La$ as large or large than in our Universe,
and it falls further the longer it takes for intelligent observers
to arise (ie, for larger $\tau$). For $\tau = 1$ we obtain a
probability of  $9 \cdot 10^{-6}$, which drops to $4\cdot10^{-12}$
for $\tau = 10$. The situation is only marginally better in the
optimistic situation where intelligent observers evolve before
one--tenth of the current age of the universe, since
$Pr(R>1\vert\tau=0.1) = 5\cdot10^{-4}$.

It is worth noting that the conclusion that low $\La$ is favored
does not depend on the observer civilization foolishly squandering
all of its resources on observing $\La$. Rather, it  requires only
that civilizations spend a fraction of their resources  doing so
which does not depend on (or at least does not decrease with)
$\La$. One might wonder why a civilization would bother
``observing'' $\La$ more than once.  First, since we have
calculated the maximum number of thermodynamic processes, we must
understand that ``observing'' should be rather broadly defined.
In particular it would include ``remembering'' the cosmological
constant (i.e.  consulting permanent records), or communicating
the value of the cosmological constant to other members of the
civilization, including one's descendants.  Thus, it would
actually be difficult for a civilization to stop ``observing'' the
cosmological constant. Secondly, there is actual motivation for
continuing to observe the rate of expansion to check to see if the
dark energy density has changed, since this alone will allow one
to take advantage of the decline in the de Sitter temperature to
prolong the civilization's existence.

\subsection{Many observers scenario}

So far, we have worked exclusively in the rare observer limit --
where each intelligent observer is free to collect all of the
energy within their apparent horizon without competition from
other observers. One might imagine that as the density of
observers rose, one would mitigate the preference for low
$\Lambda$, but the case is by no means so clear.  If the observer
density is high, then the observers will come into competition for
the universe's (or at least their Hubble volume's) same scarce
resources.   Our own historical experience is that such
competition never leads to negotiated agreement to use those
resources as conservatively as possible.  More likely is that the
competition for resources will lead to some substantial fraction
of those resources being squandered in warfare until only one of
the observers remains.  Moreover, unless they eliminate all
possible competitors, observers will continue to spend their
finite supply of energy at a rate exceeding that which would
otherwise be necessary. What is clear is that given our inability
to predict or measure either the density of intelligent observers
or the way in which they would behave when they meet, our ability
to use anthropic reasoning can only be further compromised.

Indeed, it is (not surprisingly) impossible to escape the
psychology and sociology questions when the density of observers
is high. No doubt one could perform some particular calculation of
such a scenario -- assuming, for example, that all the observers
agree to use only their local resources and not to poach on each
other, or by drawing inspiration from a game--theory approach
about competitors evolving in an environment with limited
resources. But any such assumptions would be rather strong, and it
would be effectively impossible to test.  The point is, that since
such questions inevitably intrude, one cannot do a meaningful
calculation.  But certainly the anthropic prediction will depend
on the answer to these effectively unknowable questions.

Finally, in our MANO approach we must specify what fraction of its
available resources a civilization would devote to measuring the
cosmological constant. We have used here as measure the maximum
possible number of observations, but the argument would be
unchanged if civilizations used only a fixed fraction (however
small) of that maximum. Of course a civilization would be stupid
to spend all its energy in this task, and clearly the answer one
gets in terms of the posterior pdf for $\La$ depends on the
fraction of the energy consumption of the civilization as a
function of time. We have assumed that the fraction is zero until
the ambient temperature reaches the de Sitter temperature and then
a constant (it need not be 100\% -- the probability distribution
for $\Lambda$ would remain unchanged if it were any other constant
value.) Since we are not trying to prove that MANO is {\em the}
correct way to weight universes, we make no attempt to justify
that this is {\em the} correct functional dependence of the
observation rate on time, just that it is a perfectly reasonable
dependence. Other observation strategies -- such as observe once
and never observe again -- may also be perfectly reasonable.  One
could also argue that instead of counting every observation one
should count only once each observation in a causally connected
region. But this is part of the problem -- can we ever hope to
understand, nay predict, the psychology of all intelligent
civilizations from first principles, predict which psychology will
produce the longest lasting (``dominant'') civilizations, and so
infer the distribution of observation--strategies that would
result? We fear not.

\section{Conclusions}

We have argued that anthropic reasoning suffers from the problem
that the peak of the selection function depends on the details of
what exactly one chooses to condition upon -- be it the number of
observers, the fraction of baryons in halos or the total number of
observations observers can carry out. A weighting scheme according
to the maximum number of possible observations implies that the
expected value of $\Lambda$ is logarithmically close to its
minimum allowed non--negative value (or is zero or negative),
contrary to the usual result. In its usual formulation, the
anthropic principle does not offer any motivation -- from either
fundamental particle physics or probability theory -- to prefer
one weighting scheme over another, and in particular one that does
not lead to paradoxical or self--contradictory conclusions of the
type described in~\cite{Neal}. Lacking either fundamental
motivations for the required weighting, or other testable
predictions, anthropic reasoning cannot be used to explain the
value of the cosmological constant. We expect that similar
statements apply to any conclusions that one would like to draw
from anthropic reasoning.


\begin{theacknowledgments}
RT is supported by the Royal Astronomical Society through the Sir
Norman Lockyer Fellowship. GDS is supported in part by fellowships
from the John Simon Guggenheim Memorial Foundation and Oxford's
Beecroft Institute for Particle Astrophysics and Cosmology, and by
grants from the US DoE and NASA to CWRU's particle astrophysics
theory group. RT would like to thank the organizers of the Dark
Side of the Universe 2006 conference for a very interesting
meeting and the European Network of Theoretical Astroparticle
Physics ILIAS/N6 under contract number RII3-CT-2004-506222 for
supporting his participation to this conference.
\end{theacknowledgments}




\bibliography{sample}

\IfFileExists{\jobname.bbl}{}
 {\typeout{}
  \typeout{******************************************}
  \typeout{** Please run "bibtex \jobname" to optain}
  \typeout{** the bibliography and then re-run LaTeX}
  \typeout{** twice to fix the references!}
  \typeout{******************************************}
  \typeout{}
 }

\end{document}